\documentclass{aastex}
\usepackage{emulateapj5}

\slugcomment{To appear in The Astrophysical Journal}
\shorttitle{[OIII] images of Seyferts}
\shortauthors{Schmitt et al.}
\begin{document}

\title{A Hubble Space Telescope Survey of Extended [OIII]$\lambda$5007\AA\
Emission in a Far-Infrared Selected Sample of Seyfert Galaxies:
Results\altaffilmark{1}}

\author{H. R. Schmitt\altaffilmark{2,3,4}, J. L. Donley\altaffilmark{5,6},
R. R. J. Antonucci\altaffilmark{7}, J. B. Hutchings\altaffilmark{8},
A. L. Kinney\altaffilmark{9}, J. E. Pringle\altaffilmark{10,11}}

\altaffiltext{1}{Based on observations made with the NASA/ESA Hubble Space
Telescope, which is operated by the Association of Universities for Research
in Astronomy, Inc., under NASA contract NAS5-26555.}
\altaffiltext{2}{National Radio Astronomy Observatory, 520 Edgemont Road,
Charlottesville,VA22903.}
\altaffiltext{3}{Jansky Fellow.}
\altaffiltext{4}{email:hschmitt@nrao.edu}
\altaffiltext{5}{Department of Astronomy and Astrophysics, The Pennsylvania
State University, 525 Davey Laboratory, University Park, PA16802.}
\altaffiltext{6}{National Radio Astronomy Observatory, P.O. Box 0, Socorro,
NM87801.}
\altaffiltext{7}{University of California, Santa Barbara, Physics Department,
Santa Barbara, CA 93106.}
\altaffiltext{8}{Dominion Astrophysical Observatory, Herzberg Institute of
Astrophysics, National Research Council of Canada, 5071 West Saanich Road,
Victoria, BC V9E 2E7, Canada.}
\altaffiltext{9}{NASA Headquarters, 300 E Street SW, Washington, DC20546.}
\altaffiltext{10}{Institute of Astronomy, Cambridge University, Madingley Road,
Cambridge CB3 0HA, UK.}
\altaffiltext{11}{Space Telescope Science Institute, 3700 San Martin Drive,Baltimore, MD21218.}

\begin{abstract}
We present the results of a Hubble Space Telescope (HST) survey of extended
[OIII] emission in a sample of 60 nearby Seyfert galaxies (22 Seyfert 1's
and 38 Seyfert 2's), selected by mostly isotropic properties. The comparison
between the semi major axis size of their [OIII] emitting regions
(R$_{Maj}$) shows that Seyfert 1's and Seyfert 2's have similar distributions,
which seems to contradict Unified Model predictions. We discuss possible ways
to explain this result, which could be due either to observational limitations
or the models used for the comparison with our data. We show that Seyfert 1
Narrow Line Regions (NLR's) are more circular and concentrated than
Seyfert 2's, which can be attributed to foreshortening in the former.
We find a good correlation between the NLR size and luminosity, following
the relation R$_{Maj}\propto$L([OIII])$^{0.33\pm0.04}$, which is flatter
than a previous one found for QSO's and Seyfert 2's.
We discuss possible reasons for the different results, and their
implications to photoionization models. We confirm previous results which
show that the [OIII] and radio emission are well aligned, and also find no
correlation between the orientation of the extended [OIII] emission and the
host galaxy major axis. This agrees with results showing that the torus axis
and radio jet are not aligned with the host galaxy rotation axis,
indicating that the orientation of the gas in the torus, and not the spin
of the black hole, determine the orientation of the accretion disk, and
consequently the orientation of the radio jet.

\end{abstract}


\section{Introduction}

Ground-based narrow-band imaging of Seyfert galaxies (Pogge 1988a,b,1989;
Haniff, Wilson \& Ward 1988) revealed extended emission in several of these
objects. The fact that several Seyfert 2 galaxies presented conically shaped
Narrow Line Regions (NLR's) - e.g. NGC\,1068 and NGC\,4388 - indicate that the
source of radiation ionizing the gas is collimated, a result which gave strong
support to the Unified Model. This model proposes that Seyfert 1
and Seyfert 2 galaxies are the same kind of object, a central black hole
surrounded by a torus of molecular gas and dust (Antonucci 1993, Urry \&
Padovani 1995). The identification of the AGN type
depends on whether the nucleus is observed through the torus pole, in which
case the Broad Line Region (BLR) is detected and the galaxy is classified as
a Seyfert 1, or through the torus equator, in which case the BLR is hidden
from direct view, only the NLR is detected, and the galaxy is classified as a
Seyfert 2.

   Many tests over the past 20 years have led to wide acceptance of the
Unified Model as explaining some of the differences between AGN types.
The most direct evidence in favor of the model is the detection of
polarized broad emission lines in the spectrum of Seyfert 2 galaxies
(Antonucci \& Miller 1985, Miller \& Goodrich 1990; Kay 1994; Tran 1995),
which indicates that the BLR is hidden from direct view and we see it scattered
in our direction. Another important line of evidence is the detection of high
column densities of gas absorbing the nuclear X-ray spectrum of Seyfert 2's
(Turner et al. 1997). 

Two more lines of evidence in favor of the model are the deficit
of ionizing photons in
the spectrum of Seyfert 2's (Wilson, Ward \& Haniff 1988; Storchi-Bergmann,
Mulchaey \& Wilson 1992; Schmitt, Storchi-Bergmann \& Baldwin 1994)
and the collimation of the ionizing radiation by the torus walls, discussed
above. Following the papers of Pogge (1988a,b) and Haniff et al. (1988),
several other Seyfert galaxies were found to have conically shaped NLR's
(Tadhunter \& Tsvetanov 1989; Storchi-Bergmann, Wilson \& Baldwin 1992;
Schmitt et al. 1994; among others).

The most complete study of the morphology and sizes of the NLR's of Seyfert
galaxies was done by Mulchaey, Wilson \& Tsvetanov (1996a,b). These authors
compared ground based [OIII] and H$\alpha+$[NII] images of a sample of early
type Seyferts, with models for the distribution of NLR gas and inclination of
the torus relative to the line of sight, in an attempt to determine the
intrinsic gas distribution in these sources. They found no significant
difference in the sizes and morphologies of the NLR's of Seyfert 1's and
Seyfert 2's. They also found a large number of Seyfert 2's with halo-like
NLR's, which could not be explained by their models. However, they point out
that several of the halo-like Seyfert 2 galaxies presented V-shaped
structures in ionization maps (obtained dividing the [OIII] by the H$\alpha$
images). They claim that the halo morphology was most likely due to
the low spatial resolution ($\sim$2\arcsec) of their images, which
probably was not good enough to resolve the most compact sources.
Some Seyfert 2's are known to have BLR's hidden by dust in the host
galaxy disk (Keel 1980; Lawrence \& Elvis 1982; Maiolino \& Rieke 1995)
which may also result in halo-like NLR's.

The high resolution achievable with HST is an
important tool in the study of the NLR of Seyfert galaxies. Several papers
have used HST to image a large number of sources with narrow band filters
centered at [OIII]$\lambda$5007\AA\ and H$\alpha$ (Wilson et al. 1993;
Mulchaey et al. 1994; Bower et al. 1994; Capetti et al. 1996; Schmitt \&
Kinney 1996; Ferruit et al. 2000; Falcke, Wilson \& Wilson 1998; Simpson et
al. 1997; Macchetto et al. 1994; Boksenberg et al. 1995; Cooke et al. 2000).
These papers found a large number of Seyfert 2 galaxies with extended [OIII]
emission, and also showed that the interaction between the radio jet and the
ambient gas can be an important factor in shaping the NLR, either by ionizing
it through shocks, or by compressing the gas and increasing its emissivity.

Although HST has been used to image dozens of Seyfert 2 galaxies, until
recently, only ten or so Seyfert 1's had images available in the archive,
most of which were obtained before the correction of spherical aberration. 
Using a compilation
of all the narrow band images of Seyfert galaxies available by 1995,
Schmitt \& Kinney (1996) compared the properties of the NLR of Seyfert 1's
and Seyfert 2's. Their analysis came to an intriguing conclusion, that
Seyfert 1 galaxies have much smaller NLR's than Seyfert 2's, even when
orientation effects were taken into account. However, the significance of
their result can be questioned, since their sample was extracted from the
HST archive and involved data from different projects. Wilson (1997) and
Nagar \& Wilson (1999) pointed out that, due the spherical aberration problems
during the first years of HST, several of these projects selected their
sources based on previous knowledge of extended emission, which certainly
biased the results. The fact that all the Seyfert 1's used by Schmitt \&
Kinney (1996) had only spherically aberrated observations also influenced their  
results. These images had to be deconvolved in order to correct for this effect, 
and faint extended emission around a strong point source could have been
easily washed out, making it difficult to detect diffuse extended emission
around the nuclei of Seyfert 1 galaxies.

In order to address these problems and perform an unbiased comparison between
the properties of the NLR's of Seyfert 1 and Seyfert 2 galaxies, we were 
awarded an HST snapshot project during cycle 9, to observe the extended
[OIII]$\lambda$5007\AA\ emission of the Seyfert galaxies in the 60$\mu$m sample
(de Grijp et al. 1987; 1992; Kinney et al. 2000). This survey obtained images
for 42 Seyfert galaxies,
which, combined with archival data for another 18 galaxies observed in a
similar way, created a homogeneous database of [OIII] images for a sample
of 60 Seyfert galaxies. These galaxies were
selected on mostly isotropic properties, 60$\mu$m luminosity and
25/60$\mu$m color, which ensures that we are studying Seyfert 1's and Seyfert 2's 
with similar luminosities, as well as other basic properties. 
Nevertheless, we point out that our sample
is not complete, since it does not include Seyfert galaxies with cool infrared
emission, and suggest reading Schmitt et al. (2001) for a discussion of
possible anisotropy problems in the way the sample was selected.
This sample does however mitigate many of the problems faced by previous studies, 
like selection effects, the small number of Seyfert 1's with [OIII] images, and
the lack of homogeneous observations.

A detailed description of the observations, reductions and measurements
being used in this paper is given by Schmitt et al. (2003). In that
paper we also compared some properties of the galaxies with observed [OIII]
images with those of the entire 60$\mu$m sample, and show that the observed
galaxies are an unbiased subsample, representative of the entire sample,
that can be used to draw robust statistics of the
properties of the NLR's of Seyferts. This paper analyzes the results of this
survey. In Section 2 we give a brief description of the data and measurements
being used. Section 3 presents the comparison between the
morphology and sizes of the NLR's of Seyfert 1's and Seyfert 2's, and a
comparison with the predictions from Mulchaey et al. (1996b) models.
In Section 4 we study the correlation between the size and luminosity of
the NLR, while in Section 5 we discuss the correlation between the
radio and narrow-line emission. A summary of the results is presented in
Section 6.

\section{Data and Measurements}

The data used in this Paper were obtained with HST. Our sample is composed
of 60 Seyfert galaxies (22 Seyfert 1's and 38 Seyfert 2's) selected based
on their far infrared properties (Kinney et al. 2000). The observations for
58 of the 60 galaxies were obtained with the WFPC2 camera, one Seyfert 2
galaxy was observed with the FOC camera and another one with the WF/PC1 camera.
The on-band [OIII] observations for 56 galaxies were done with the Linear
Ramp Filter, while the remaining ones were done with regular narrow-band
filters. Typical exposure times of these images were between 10~min and
20~min, split into 2 or more exposures to allow easier elimination of
cosmic rays. Short continuum observations were also obtained, to allow the
subtraction of the host galaxy contribution to the on-band image. These images
were obtained with filter F547M, and, in some cases, with the Linear Ramp
Filter, having typical integration times between 1~min and 2~min. In most
cases the continuum images were not split into multiple exposures, and required
some extra work to eliminate cosmic rays.

The reductions were done by the HST pipeline, with the exception of the
flat field correction of the images obtained with the Linear Ramp filter.
Since there are no flat fields for this filter, we used the one for
F502N, which has a similar wavelength and bandwidth. The images were
registered, combined, cleaned of cosmic rays and background subtracted.
The calibration of the continuum images was done using the information on
their headers, while for those obtained with the Linear Ramp Filter we
used the WFPC2 Exposure Time Calculator for extended sources, which gives
an accuracy of $\sim$5\%. Finally,
continuum-free [OIII] images were created by subtracting continuum images,
scaled by the bandwidth of the on-band images, from the on-band images.

The continuum-free [OIII] images were clipped at the value of 3$\sigma$,
and the [OIII] fluxes were measured by integrating the flux inside a
rectangular region which included all the visible emission
(F([OIII])$_{int}$). A comparison between these values and values obtained
from the literature (Schmitt et al. 2003) shows a fairly good
agreement, with a scatter of $\sim$0.25~dex. We measured the nuclear [OIII]
fluxes of these galaxies (F([OIII])$_{nuc}$), obtained integrating the flux
inside a circular aperture with radius of 100 pc, centered at the nucleus.
These images were also used to measure the dimensions of the NLR's, like the
effective radius, semi major and semi minor axis of the [OIII] emission
(R$_e$, R$_{Maj}$ and R$_{Min}$), the position angle of the [OIII] major axis
and the offset between the nucleus and the centroid of the [OIII] emitting
region. We assume H$_0=75$~km~s$^{-1}$~Mpc$^{-1}$ throughout this paper.
The results of the statistical tests performed in this paper
are summarized in Table 1, where we give the property being compared,
the distributions being compared, the KS test probability, and the Figures
where the data are presented.

\section{Sizes and Morphologies of the NLR's}

The morphologies and sizes of the NLR's of Seyferts are important diagnostics
for the distribution of the gas in the nuclear region of these galaxies. 
Narrow-band images have also been used, as pointed out above, to show that the
nuclear radiation, which ionizes this gas, is collimated or beamed.
The idealized picture of the Unified Model suggests that the NLR
of Seyfert 1's, where the torus is seen pole-on, should have a halo like
morphology, while Seyfert 2's, where the torus is seen through the equator,
should have conically shaped NLR's. However, even from low-resolution
ground-based observations it was clear that this simple model needed some
improvements. One example which clearly shows this is
the detection of a conically shaped NLR in the Seyfert 1 galaxy NGC4151
(Perez et al. 1989; Pogge 1989).

A more detailed comparison between the morphology of the [OIII] images of
Seyferts with geometrical models for the shape of the NLR was presented by
Mulchaey et al. (1996b). These authors modeled the geometry of the NLR
assuming a torus with semi opening angle of 35$^{\circ}$, and a central engine
ionizing gas distributed in two different geometries, a sphere and a disk.
They showed that the simplest expectation, described above, is true only
when the NLR has a spherical distribution of gas with properties suitable
for producing the narrow emission lines when illuminated. In the case when
the gas lies in a disk, possibly the host galaxy disk, both Seyfert
types can have conically shaped NLR's. One of the principal predictions
from their models is that the NLR's of Seyfert 1's, seen in projection,
should on average be smaller than Seyfert 2's.

\subsection{NLR sizes}

In Figure 1 we present the cumulative distribution of the logarithm of R$_{Maj}$.
A visual inspection of this Figure indicates that there is no difference 
in the distribution of values for Seyfert 1's and Seyfert 2's. The KS test
gives a 90\% chance that two samples drawn from the same parent population
would differ this much, thus the two distributions are very similar (in fact
fortuitously so). Comparing these results with the models from Mulchaey
et al. (1996b), we find that the observed distribution seems to agree better
with that where the NLR gas is in the host galaxy disk. The fact that several
Seyfert 1 galaxies present very extended NLR's (e.g. MRK\,79, MCG+08-11-11)
gives support to this interpretation. However, we point out that several
Seyfert galaxies, including some in our sample, are known to have outflows,
indicating that the gas does not necessarily have to be confined to a disk.
Some of these galaxies are NGC\,4388 (Corbin et al. 1988; Veilleux et al.
1999), NGC\,3516 (Mulchaey et,al. 1992), NGC\,3281 (Storchi-Bergmann et al.
1992), NGC\,1068 (Crenshaw \& Kraemer 2000; Cecil et al. 2002),
NGC\,4151 (Hutchings et al. 1998; Crenshaw et al. 2000). Furthermore,
Antonucci (2002) showed that the [OIII] emission in Seyfert 1 galaxies is
blueshifted with respect to the host galaxy stars, while Seyfert 2's have a
symmetric distribution around the systemic velocity, suggesting that outflows
may be common.

One problem that is clearly apparent in the comparison of the observed
results with the models is the similarity between the distribution of
NLR sizes of Seyfert 1's and Seyfert 2's. However, this result might
be explained by several possible reasons, which we outline here. 
Since corrections for these effects could be arbitrary and not very
easy to performed, we decided to just list them and do not try to
do anything with the data.

It is possible that several galaxies, as many as 5 Seyfert 1's and
1 Seyfert 2 (e.g. MRK\,705, NGC\,5548, NGC\,7213) have unresolved [OIII]
emission, even at the
resolution of HST. The WFPC2 manual shows that at a wavelength around
5000\AA\, the flux of a point spread function is reduced to $\sim$1\% of
the peak flux at a radius of 0.5\arcsec. The inspection of the several [OIII]
images of the galaxies described above (Schmitt et al. 2003), shows only
a compact nuclear sources. The [OIII] flux of these galaxies falls to
approximately 1\% of the peak flux at distances of $\sim$0.5\arcsec\
away from the nucleus, which is consistent with the proposed interpretation.

Among the resolved images, as many as 6
Seyfert 2's and 2 Seyfert 1's (e.g. NGC\,3281, NGC\,4388), may have their NLR
sizes reduced by shading from the torus or host galaxy disk. 
That is, in some galaxies we observe only the side of the NLR which is
facing us, while the other side is blocked by the
host galaxy disk. This can reduce the observed NLR size to approximately
half of the intrinsic one. 

We tried to estimate how important the effect of hiding part of the conical
NLR by the galaxy disk might be in the determination of their sizes, and in
particular, if this effect is more important in Seyfert 2's compared to
Seyfert 1's. In Figure 2 we present the Offsets between the peak of the
continuum and the centroid of the [OIII]-emitting region. If the
NLR is centered around the nucleus of the galaxy we should obtain
Offset$=0$, while in the case where one side of the NLR is hidden by
the galaxy disk we should obtain larger Offsets, with the maximum value
being 1. An inspection of Figure 2 shows that most of the galaxies in our
sample have NLR's which are more or less centered around the nucleus,
with Offset values smaller than 0.2. Seyfert 1 galaxies seem to be concentrated
towards values smaller than 0.4, while Seyfert 2's present a tail going
up to 1. However, the KS test does not show a statistically significant
difference between the distribution of values for Seyfert 1's and Seyfert 2's,
with a 33\% probability that two samples drawn from the same parent population
would differ this much. This result indicates that this effect can influence
the determination of the sizes of NLR's, but is not the only factor, or even
the most important one, causing both Seyfert types to have similar
distributions of R$_{Maj}$.

Orientation effects may also play an important role in the observed size of
the NLR of Seyfert 1's, by enhancing their surface brightnesses. Since their
conical NLR's are foreshortened by the fact that they are observed end-on, this
means that their NLR emission is integrated along a larger path length,
resulting in enhanced [OIII] surface brightnesses at all scales. The NLR's
of Seyfert 2 galaxies are observed edge-on, corresponding to smaller
path lengths and lower surface brightnesses in regions which are far away
from the nucleus. As a result of this effect, it is easier to detect extended
emission in Seyfert 1's, relative to Seyfert 2's, especially in a
surface brightness limited survey like ours.

Other important factors that should be taken into account when
comparing the observed distribution to Mulchaey et al. (1996b) models,
is the fact that these models do not take into account different source
luminosities, or different torus opening angles. We will show in the next
section that the size of the NLR increases with the luminosity of the
AGN, so this size can vary significantly in a sample with a large range
of luminosities like ours. The fact that the torus half opening angle can
be different from that used in their models (35$^{\circ}$), can also change
the predicted distribution of NLR sizes for Seyfert 1's and Seyfert 2's.
In reality one should not expect the torus opening angle to be constant
in a sample with such a wide range of luminosities, and  could in fact
depend on the luminosity of the source (Lawrence 1991). Just for illustrative
purposes, consider the torus half opening angle of 48$^{\circ}$ that
Schmitt et al. (2001) found comparing the relative number of Seyfert 1's
and Seyfert 2's in the parent population of galaxies used in this paper.
Such a torus will generate an ionization cone where the size of the base
is 1.5 times the size of the side of the cone. When compared to the base
of the cone generated by a torus opening angle used in Mulchaey's models, 
we find that the base of the cone of these models is 25\% smaller than that
of a cone with half opening angle of 48$^{\circ}$.
Since the base of the cone indicates how large a conical NLR would be
if the torus is observed pole-on, like in Seyfert 1's, this
indicates that these galaxies could have NLR's with sizes similar to those of
Seyfert 2's. Furthermore, if there is a range of torus opening angles,
in a sample like the one used here, which was randomly drawn from a complete
sample, the average opening angle of Seyfert 1's will be larger than
that of Seyfert 2's. This is due to the fact that galaxies with larger torus
opening angles are, on average, more likely to be seen as Seyfert 1's.
If all the factors described here were taken into account in the models, they
would result in different distribution of NLR sizes for Seyfert 1's and
Seyfert 2's, which could look much more like ours.

\subsection{NLR Morphologies}

Besides the sizes of the NLR regions, we also compared other
morphological measures. A simple picture of the Unified Model, 
predicts that since the conical NLR of Seyfert 1's
is seen close to end-on, they should look rounder than the NLR of
Seyfert 2's, which are seen closer to edge-on. Second, due to this projection
effect, we also expect the NLR's to be more concentrated towards the
nucleus in Seyfert 1's than in Seyfert 2's, since we are integrating a larger
column density of gas along the line of sight in those galaxies (if the NLR is
more or less uniformly filled rather than hollow).

We first compare the ratio between the apparent semi major and semi minor axes
of the [OIII] emission. Figure 3 shows the R$_{Maj}$/R$_{Min}$ distribution,
which is peaked at values smaller than 1.5, and decreases to higher values.
Most of the Seyfert 1's in the sample are concentrated in this bin, while
Seyfert 2's have a broader distribution, extending to much higher values.
We find that 68\% of the Seyfert 1 galaxies have R$_{Maj}$/R$_{Min}<1.5$,
indicating that their NLR's have more of a halo like morphology. On the
other hand, 68\% of the Seyfert 2 galaxies have R$_{Maj}$/R$_{Min}>1.5$,
which indicates that their NLR's are more elongated.

Comparing the distribution of R$_{Maj}$/R$_{Min}$ values of Seyfert 1's and
Seyfert 2's using the KS
test, we find that there is only a 2.3\% probability that two samples
drawn from the same parent population would differ this much. Even if we
eliminate the two Seyfert 2 galaxies with the largest ratios, we still find
a significant difference between the two samples, with the KS test giving
only a 3.4\% probability. These results are consistent with the scenario where
the conical NLR if Seyfert 1 galaxies are seen closer to end-on, while 
Seyfert 2's are seen closer to edge-on.

The compactness of the NLR emission can be tested in several ways, four
of which are presented here. We first compare the ratio of the nuclear
[OIII] flux, which corresponds to the flux inside an aperture with a radius
of 100 parsecs, centered at the nucleus, to the integrated [OIII] flux.
The left panel of Figure 4 shows that the F([OIII])$_{nuc}$/F([OIII])$_{int}$
distribution, where we can see that Seyfert 1's are skewed towards higher
values and Seyfert 2's are skewed towards lower ones, indicating that the
emission is more concentrated towards the nucleus in the former. Comparing
the two distributions using the KS test we find that there is only a 0.47\%
chance that two samples drawn from the same parent population would differ this much.

A similar test for the concentration of the [OIII] emission around the nucleus
consists in comparing the nuclear [OIII] luminosities, which we present
in the right panel of Figure 4. As for the F([OIII])$_{nuc}$/F([OIII])$_{int}$
ratio, we find that the two Seyfert types have significantly different
distributions, with the Seyfert 1's having higher L([OIII])$_{nuc}$'s.
Since both types have similar L([OIII])$_{int}$ distributions (Schmitt et al.
2003), we conclude
that the emission is more concentrated towards the nucleus in Seyfert 1's.
Comparing the L([OIII])$_{nuc}$ distributions using the KS test, we find
that there is only a 0.14\% chance that two samples drawn from the same parent
population would differ this much.

Two alternative ways to confirm that the NLR emission is more concentrated
towards the nucleus in Seyfert 1's than in Seyfert 2's, is to compare the
distribution of their effective radii, and the ratios between the effective
radius to the semi major axis of the NLR. In the left panel of Figure 5 we
present the distribution of the logarithm of the effective radii of the NLR's
of Seyfert 1's and Seyfert 2's. As expected, from the results obtained in
Figure 4, Seyfert 1's have on average smaller R$_e$'s than Seyfert 2's,
with the KS test finding only a 0.5\% probability that two samples drawn
from the same parent population would differ this much. The R$_e$/R$_{Maj}$
distribution (Figure 5 right panel) confirms these results, with
Seyfert 1's presenting smaller values than Seyfert 2's. The KS test shows that
two samples drawn from the same parent population would differ as much as
these two only 0.08\% of the time.

\section{NLR Size-Luminosity Relation}

An important question in the study of AGN's is how the size of the
NLR scales with the luminosity of the central source. If such a correlation
exists over a large range of luminosities (as in our sample), we can
use this to improve our knowledge about the ionization structure
of the NLR, as well as derive an extra input in photoionization models.

So far, most of the work on this subject has been concentrated on the
size-luminosity relation of the Broad Line Regions (BLR's), which
has an important application in the determination of black hole masses.
Reverberation mapping studies have been used to determine the slope of this
relation in Seyfert 1's and QSO's, where the typical values found are
R$_{BLR}\propto$L$_{Cont}^{0.5}$ (e.g. Peterson et al. 2002) and
R$_{BLR}\propto$L$_{Cont}^{0.7}$ (Kaspi et al. 2000), respectively. However,
little work has been done in the case of NLR sizes. Mulchaey et al. (1996b)
found, based on their ground based observations, that R$_{Maj}$ and
[OIII] luminosity are correlated, but considered that this result could very
well have been due to a selection effect. Pogge et al. (2000) found that
the NLR's of LINER's are significantly smaller than those of Seyferts,
consistent with the fact that they are lower luminosity AGN's.
More recently, Bennert et al. (2002), repeated the analysis of Mulchaey
et al. (1996b) for a sample of 7 QSO's and 7 Seyfert 2
galaxies. They found a good correlation between these two quantities, with
a slope of approximately 0.5. They discuss the implication of this result
and the fact that this slope is similar to that found for BLR's, but caution
that their result is based on a small number of galaxies and requires further
investigation, using a larger sample.

\subsection{Seyfert Galaxies}

The sample and measurements being used in this paper are ideal to determine
if there is a correlation of NLR size and luminosity in Seyferts.
If one assumes that the [OIII] luminosity can
be used as a surrogate for the nuclear continuum luminosity, as indicated by
the correlations between emission line and continuum strength found by several
authors (Yee 1980; Shuder 1981; Cid Fernandes et al. 2001; Ho \& Peng 2001),
one gets that our sample spans about 3 decades in nuclear luminosity
(39.2$\leq$ log L([OIII])$\leq$42.2). Another advantage of our sample is the
fact that the galaxies are nearby, so we are able to resolve the NLR emission.

One could argue that it would be more appropriate to do such measurements
using H$\alpha$ or other recombination line images, since [OIII] is a
collisionally excited line. However, Falcke et al. (1998) and Ferruit et
al. (2000) showed that the H$\alpha$ and [OIII] images of Seyfert galaxies
have similar structures and extents, indicating that using [OIII] is not a
problem. Furthermore, H$\alpha$ has the major drawback of being highly
sensitive to star formation. Using this line would increase the chance
of contaminating the measurements with HII regions.

We present in Figure 6 a series of log R$_{Maj} \times$ log L([OIII]) plots.
The left panel presents all the Seyferts in the sample, where
we can clearly see a correlation between these two measurements. The Spearman
rank test gives a probability smaller than 0.01\% that a correlation is not
present. This plot also presents a linear fit to the data, which gives the
following relation:

$$log R_{Maj} = (0.33 \pm 0.04) \times log L([OIII]) -10.78 \pm 1.70$$

\noindent
with a correlation coefficient of 0.627.

A conservative estimate of the uncertainty in the measurement of R$_{Maj}$
suggests that these values can be off by $\sim$2 pixels (0.2\arcsec) in the
worst cases. On average, this corresponds to an error smaller than 10\%,
so we decided to fit the data using uniform weighting of the points.

Figure 6 also presents the same plot, separating the galaxies into
Seyfert 2's and Seyfert 1's only (middle and right panel, respectively).
We can see that the correlation is still present, although with some small
differences. The best fitting line for Seyfert 2's is:

$$log R_{Maj} = (0.31 \pm 0.04) \times log L([OIII]) -10.08 \pm 1.80$$

\noindent
with a correlation coefficient of 0.677, and the Spearman rank test showing
a probability smaller than 0.01\% that a correlation is not present.
For Seyfert 1's, we find that the best fitting line is:

$$log R_{Maj} = (0.41 \pm 0.08) \times log L([OIII]) -14.06 \pm 3.45$$

\noindent
with a correlation coefficient of 0.666, and a Spearman rank test probability
of 0.13\% that a correlation is not present.

A comparison between the best fitting line for the entire sample and that
obtained using only Seyfert 2's, shows a very small difference in all
the parameters, as well as in the correlation coefficient. On the other hand,
when we compare the results obtained for Seyfert 1's with those obtained
for Seyfert 2's or the entire sample, we find that Seyfert 1's
have a steeper slope, with a larger uncertainty. Considering this
uncertainty, we find that this slope is not significantly different from
the other two.

The scatter in the correlation for Seyfert 1's is larger than that
for Seyfert 2's, which can be due to
projection effects. Since the NLR of those galaxies is observed closer to
end-on, this can result, in some cases, in smaller dimensions than the
intrinsic ones, increasing the scatter in the plot. This problem is reduced
in Seyfert 2's, because their NLR's are seen closer to edge-on, resulting
in dimensions closer to the intrinsic ones.

We considered whether the observed correlations could be due to the fact
that all our observation had similar detection limits, which corresponded
to a more or less constant [OIII] surface brightness. In a situation like
this, the luminosity of the NLR increases as a function of its area, and
one would expect a relation of the form R$_{Maj}\propto$ L([OIII])$^{0.5}$.
This relation is much steeper than the observed one, and inconsistent with
the measurements, indicating that the observational result is not dominated by
this effect.

A problem similar to this, which could influence our observation, would be the
difficulty to detect the fainter, most extended emission in the more distant
galaxies, due to the uniform detection limit of our observations. The most
distant galaxies usually are the more luminous ones in a flux limited sample,
and, since the fainter gas may not necessarily make a significant contribution
to the integrated [OIII] luminosity, this would flatten the observed
correlation. We tested if our results were due to this effect by dividing the
sample in half, based on their distances, and fitting lines to this data. For
those galaxies with radial velocities smaller than 5300 km~s$^{-1}$ we get that
R$_{Maj}$ goes with the 0.30$\pm$0.06 power of L([OIII]), and for the most
distant half of the sample we get a relation with the 0.32$\pm$0.05
power. We do not differentiate between Seyfert 1's and Seyfert 2's in this
analysis because there are not enough galaxies to allow us to do so. Again,
we find no significant difference in the slopes of the two subsamples of
Seyferts, indicating that the observed result should not be due to this
distance-size effect.

As a last test, instead of using the integrated values, we use the
effective radii and luminosities. In Figure 7 we present the
log~R$_e\times$log~L([OIII])$_e$ plot, where we can see that a correlation
still is present, but with a much larger scatter and flatter slope
than that obtained using the integrated values.
The best fitting line is:

$$log R_e = (0.21 \pm 0.07) \times log L([OIII])_e -6.45 \pm 2.87$$

\noindent
with a correlation coefficient of 0.338 and a Spearman rank probability of
0.94\% that a correlation is not present.

The results presented above show that there is a correlation of the form
R$_{Maj}\propto$ L([OIII])$^{0.33\pm0.04}$ between the size and luminosity of
the NLR of Seyfert galaxies. A simple interpretation of this result suggests
that these NLR's can be represented by one of the simplest ionization
structures, a cloud of gas ionized by a central source (Str\"omgren sphere).
Albeit the geometry of an AGN is consistent with this picture, observing
such a correlation is a little surprising. In principle, the relation
R$\propto$ L$^{0.33}$ is valid only in the case of constant density, but
more elaborate and complete models may also be able to reproduce
this relation. Our observations, as well as spectroscopic
ones, show that the gas is not homogeneously distributed inside the NLR,
and there is also some evidence indicating that the gas density may
increase towards the nucleus. Interactions between radio jets and the NLR
gas can also significantly alter its shape, and possibly generate shocks,
which will represent a second source of ionization to the gas. A combination
of all these factors could in practice result in slopes steeper than the
observed one.

Our result also differs from claims that the ionization parameter does not
vary by a significant amount inside the NLR, which should result in a 
size-luminosity relation like the one found by Bennert et al. (2002)
(R$_{Maj}\propto$ L$^{0.5}$). These claims are based on the fact that emission
line ratios are uniform among different objects (see Dopita et al. 2002 for
a discussion on this subject). However, spatially resolved observations
of several Seyfert galaxies (e.g. Storchi-Bergmann et al. 1992; Fraquelli,
Storchi-Bergmann \& Binette 2000) show that the excitation of the NLR
gas decreases with distance from the nucleus. Furthermore,
Ferguson et al. (1997), using photoionization models which use locally
optimally emitting clouds, showed that given the right ensemble of clouds,
the ionization parameter can vary by a factor of more than 10 while 
emission line ratios like [OIII]/H$\beta$ and [OIII]/H$\alpha+$[NII],
which are used as excitation indicators, remain constant. All these results
indicate that the ionization parameter of the gas in the NLR falls with
distance from the nucleus, consistent with our results.

\subsection{Comparison of Seyferts and QSO's}

Compared with the results obtained by Bennert et al. (2002) using Seyfert 2's
and QSO's, our data indicate a significantly flatter slope for the
size-luminosity relation. Since their
results were based on a much smaller number of sources, we combined their
QSO measurements, converted from H$_0=65$~km~s$^{-1}$~Mpc$^{-1}$
to 75~km~s$^{-1}$~Mpc$^{-1}$, with those for our Seyfert galaxies.
These results are presented in Figure 8, where we can see that,
compared to Seyferts, the QSO points deviate from the general trend
defined by those galaxies, having R$_{Maj}$ values larger than expected.
The best fitting line to the combined  data points gives:

$$log R_{Maj} = (0.42 \pm 0.03) \times log L[OIII] -14.72 \pm 1.24$$

\noindent
with a correlation coefficient of 0.819, and the Spearman rank test showing
a probability smaller than 0.01\% that a correlation is not present.

This slope value is significantly steeper than that found using only Seyfert
galaxies. Currently we are not sure of the reason for such a difference.
One possible explanation would be that the NLR's of Seyferts and QSO's are
intrinsically different. However, given the fact that Seyferts seem to be
lower luminosity cousins of QSO's, this explanation seems to be very
unlikely. The only remaining possibility for the difference is
that it is due to some effect in the measurements of either Seyferts or QSO's.

One possibility would be the reduction technique used for some
of the QSO's presented by Bennert et al. (2002), which could have increased
their NLR sizes. For 4 of their QSO's, instead of subtracting a continuum
image from their on-band images, they subtracted only the PSF of a star.
This procedure may leave a residual host galaxy continuum contribution to
the final image, causing an overestimation of the NLR sizes and luminosities.
Nevertheless, these authors claim that their results were not influenced
by this effect, as the comparison of the images obtained in this way with
those obtained subtracting a scaled broad band image do not show a
significant difference.

A way to significantly increase the sizes of QSO NLR's would be to
include star-forming regions in the measurement. Canalizo \& Stockton
(2000a,b; 2001) showed several QSO's with circumnuclear star forming regions,
suggesting that this may be a common phenomenon. Since
Seyfert galaxies also are known to have circumnuclear starbursts
(Heckman et al. 1997; Cid Fernandes, Storchi-Bergmann \& Schmitt 1998;
Schmitt, Storchi-Bergmann \& Cid Fernandes 1999; Gonz\'alez Delgado, Heckman
\& Leitherer 2001; Cid Fernandes et al. 2001),
this effect should also be important in the determination of their
NLR sizes. However, since the QSO sample is approximately 10 times
more distant than the Seyfert sample, this effect is much more pronounced
for those galaxies, because of the angular scale of their images (usually
larger than 1~kpc per arcsec). Since both Seyferts and QSOs were observed
with the Linear Ramp Filter, which has a field of view of 13\arcsec,
most of these regions would not be seen in the images of Seyferts.
It is also possible that, since several QSO's are related to interacting
systems, presenting tidal tails and bridges (Stockton \& MacKenty 1987;
Bahcall, Kirhakos \& Schneider 1995; Bahcall et al. 1997; Dunlop et al. 2003),
that these interactions can provide debries at large distances from the
nucleus, which can be ionized by the central engine, increasing the size
of the NLR.

\section{Comparison of Position Angles}

\subsection{[OIII] versus Radio Jets}

Ground-based observations of Seyfert galaxies with large extended NLR's
have shown that there is a good correlation between the orientation of
the [OIII] emission and that of the radio jet (Wilson \& Tsvetanov 1994;
Nagar et al. 1999). These results were later confirmed by HST ones (Capetti
et al. 1996; Falcke et al. 1998). Combining this information with the fact
that Seyfert galaxies with luminous radio sources have systematically broader
[OIII] emission lines (Whittle 1992), indicates that the NLR gas can be
significantly disturbed by the interaction with the jet. This interaction
can either ionize the gas through shocks, or simply compress it, resulting
in regions of enhanced emission.

In Figure 9 we present the cumulative distribution of the difference between the
position angle of the [OIII] major axis (P.A.$_{[OIII]}$) and the P.A.
of the radio jet (P.A.$_{RAD}$). The information about the radio jet
axis was obtained from Kinney et al. (2000). We find that this
distribution is peaked around 0$^{\circ}$, with a small tail towards
higher values. The KS test shows that there is only a 0.01\% probability of
the observed distribution being drawn from a uniform distribution in the
range 0$^{\circ}$ to 90$^{\circ}$. We did not try to compare the distribution
of Seyfert 1's and Seyfert 2's, because there are only 7 Seyfert 1 galaxies
with both radio and [OIII] extended emission in our sample. Nevertheless,
an inspection of the left panel of Figure 9 provides no conclusive
evidence that both types have different distributions.

The fact that not all galaxies have perfectly aligned [OIII] and radio
emission, can in part be explained by the uncertainties in the measurement
of the P.A.'s. This effect is very important for those galaxies with
P.A.$_{[OIII]}-$P.A.$_{RAD}>30^{\circ}$, all of which have faint and only
slightly resolved radio emission, or, in the case of UGC\,2514, have [OIII]
emission extending over more than one direction. Another possibility for those
galaxies with a small misalignment is a projection effect, like the one
observed in NGC\,4151 (Pedlar et al. 1993). This effect is caused by the fact
that the torus axis is misaligned relative to the host galaxy axis. In this way
the AGN ionizes the gas in the disk of the galaxy, while the radio jet,
which is aligned to the torus axis, is at an intermediate angle between the
disk and the host galaxy axis. When seen in projection, this causes a
misalignment between the radio and [OIII] emission, which in the case of
NGC\,4151 is of the order of 30$^{\circ}$.

\subsection{[OIII] versus Host Galaxy Major Axis}

As pointed out by Clarke, Kinney \& Pringle (1998), symmetry assumptions
suggest that the accretion disk of Seyfert galaxies should be aligned
with their host galaxy disk. Consequently, one would expect all radio
jets to be perpendicular to their host galaxy major axes. However, these
predictions are clearly contradicted by the observations (Ulvestad \&
Wilson 1984; Schmitt et al. 1997; Nagar \& Wilson 1999; Pringle et al.
1999; Kinney et al. 2000) which show that the observed distribution of
jets is consistent with a homogeneous distribution in three dimensions.
Taking into consideration the relatively good alignment between the radio
and [OIII] emission, we use the measurements of P.A.$_{[OIII]}$ to verify
if the NLR of these galaxies have any prefered alignment relative to
the host galaxy disk.

In Figure 10 we present the distribution of the difference between
P.A.$_{[OIII]}$ and P.A. of the host galaxy major axis (P.A.$_{MA}$).
We point out that in the cases where the [OIII] emission had a conically
shaped NLR, but the axis of the cone was shorter than the major extent
of the NLR, we used the P.A. of the cone axis. We find that the
distribution of values for all galaxies is consistent with a uniform
distribution. The KS test gives a 56\% probability that a sample drawn
from a uniform distribution in the range 0$^{\circ}$ to 90$^{\circ}$
would differ as much as the observed one does. A comparison between the
distribution of Seyfert 1's and Seyfert 2's shows that the former are more
concentrated towards smaller values ([OIII] emission along the major axis)
while the latter seems to have a higher percentage of sources with [OIII]
emission aligned close to perpendicular to the host galaxy major axis.
However, this difference does not have a high statistical significance.
The KS test gives a 6.2\% probability that two samples drawn from the
same parent population would differ as much as the two Seyfert types.

The fact that the direction of the [OIII] emission of Seyfert galaxies
is randomly oriented relative to the host galaxy major axis, confirms
the results obtained from the comparison between the orientation of
the radio and host galaxy major axes (Kinney et al. 2000).
These results have important implications for the mechanisms responsible
for the misalignment of the accretion disks relative to the host galaxy disks.
Since the radio and [OIII] axes are reasonably well aligned, this implies that
the torus and accretion disk axes are well aligned. This is evidence against
the spin of the black hole being the mechanism responsible for changing the
orientation of the accretion disk, through the Bardeen-Petterson effect. This
effect may work at the accretion disk level, but is not important at the
distance of the torus. Instead, this result supports the idea that the
accretion disk is fed by gas from the torus, which determines its orientation,
and the orientation of the torus is determined by the origin of its gas, which
can be either internal or external from the galaxy.

\section{Summary and Conclusions}

This paper discussed the results of an HST narrow band imaging
survey of [OIII]$\lambda$5007\AA\ emission in a sample of 22 Seyfert 1 and
38 Seyfert 2 galaxies. The data, reductions and measurements used here
were presented in Schmitt et al. (2003), where we also show that
the sample of galaxies for which we have [OIII] images is a representative
subsample of the 60$\mu$m sample of Seyferts described by Kinney et al. (2000)
and Schmitt et al. (2001). This indicates that this sample can be used to
draw robust statistical results about the Unified Model. 

Our results can be divided into three groups: those which investigate the
properties of the NLR of Seyfert 1's and Seyfert 2's and compare them to
the predictions from the Unified Model; those which investigate the
relation between the size and luminosity of the NLR; and those which study
the orientation of the NLR relative to radio jets and host galaxy major axes.

In the study of NLR properties, we found the following results from the 
comparison between the measurements of Seyfert 1's and Seyfert 2's with
the predictions from the Unified Model. First, both Seyfert types have
similar distributions of NLR sizes. Compared to NLR models presented by
Mulchaey et al. (1996b), this suggests that the NLR gas is consistent with
a disk like distribution. However, several galaxies in the sample are
known to have outflows, which indicates that the disk morphology does not
apply to all galaxies. One problem that appears from this comparison
is the lack of a significant difference in the R$_{Maj}$ distribution
of the two Seyfert types. According to Mulchaey et al. (1996b), one would
expect Seyfert 1's to have, on average, smaller NLR sizes than Seyfert 2's,
but under a more detailed scrutiny, we find effects which might significantly
alter both the measurements and models.

Some of the effects we think may significantly alter the measured NLR sizes
are the possible overestimation of the NLR size in several Seyfert 1 galaxies,
which may not be significantly resolved even at the HST resolution, and the
possible underestimation of the NLR size in several Seyfert 2 galaxies, which
may have up to half of their NLR's hidden by the torus and the host galaxy
disk. Enhanced surface brightness in Seyfert 1's, due to foreshortening
of the conical NLR can also play an important role in the detection of extended
emission in these galaxies. We do not find a significant difference in the offset
between the nucleus and the centroid of the NLR in Seyfert 1's or Seyfert 2's,
suggesting that, although this problem is present in some sources, it is
not the most important factor driving their similar distribution of NLR sizes.

We also consider the possibility that Mulchaey et al. (1996b) models
could be significantly changed if one modifies their input parameters,
possibly resulting in similar NLR size distributions for Seyfert 1's and
Seyfert 2's. Their models assumed that
all galaxies have similar NLR's, and the only parameters that vary are the
distribution of gas and the inclination of the putative torus relative to
the line of sight. Since the NLR size, and probably even the torus opening
angle, depend on the luminosity of the AGN, if these effects are taken
into account in the models, one could obtain theoretical distributions
of R$_{Maj}$'s different from the ones obtained by these authors,
which could in principle be more similar to the observed one.

Following on this subject, we performed a set of tests aimed at
comparing the morphology of the two Seyfert types and the degree of
concentration of the NLR emission around the nucleus. These tests showed that
there is a higher percentage of Seyfert 1's with halo like NLR's
(R$_{Maj}$/R$_{Min}<$1.5), while those of Seyfert 2's are more elongated.
Another interesting result is the fact that Seyfert 1 NLR's are more
concentrated towards the nucleus, indicated by the fact that they, on average,
have higher F([OIII])$_{nuc}$/F([OIII])$_{int}$ ratios and  L([OIII])$_{nuc}$
than Seyfert 2's. Seyfert 1 galaxies also have smaller R$_e$ and R$_e$/R$_{Maj}$
values than Seyfert 2's, consistent with their emission being more
concentrated towards the nucleus. These results are in good agreement
with the Unified Model prediction that the conical NLR of Seyfert 1's is observed
closer to end-on, while that of Seyfert 2's is closer to edge-on.

The fact that the NLR emission is more concentrated towards the nucleus
in Seyfert 1's, explains why Schmitt \& Kinney (1996) observed that their
Seyfert 1's had much smaller NLR's than the Seyfert 2's.
All the Seyfert 1 galaxies in their sample were observed by HST with
spherical aberration. Since the
NLR emission of these galaxies is concentrated around the nucleus, the
spherical aberration washed out most of the extended emission, which could
not be recovered even after the deconvolution of the images.

The second group of results shows that there is a relation between the size
and the luminosity of the NLR's of Seyferts, with the form
R$_{Maj}\propto$ L([OIII])$^{0.33\pm0.04}$.
We tested whether there was any distance or observational bias which
could influence this result, but could not find anything. This result
indicates that the NLR follows the simple predictions of a distribution
of gas being ionized by a central source. We point out that this 
correlation can be generated by more complex models, and discuss some ways
in which it could be changed. For instance, if the density of the NLR gas
decreases with radius, this would result in a relation steeper than the
observed one. This result also agrees with those obtained from photoionization
models, which indicate that single zone models are not an appropriate
representation for NLR's (Dopita et al. 2002; Binette, Wilson \&
Storchi-Bergmann 1996). The slope obtained by us is different
from the one obtained by Bennert et al. (2002) for a sample of 7 QSO's
and 7 Seyfert 2's. These authors found a relation with a slope $\sim$0.5.
We explore possible reasons why the two relations differ,
like the undersubtraction of the host galaxy continuum, or the inclusion
of circumnuclear star forming regions in their measurements.

The last set of results obtained in this paper were based on the comparison
of the P.A. of the extended [OIII] emission to the P.A. of radio jets and
host galaxy major axes. We found a very good alignment between the radio and
[OIII] emission, confirming results from previous papers (Taylor et al. 1989;
Wilson \& Tsvetanov 1994; Capetti et al. 1996, among others). The good
correlation between these two properties was known for a long time, indicating
that the jet can significantly disturb the NLR gas, compress it, making it
radiate strongly, or even ionize it through shocks (Taylor, Dyson \& Axon 1992;
Dopita \& Sutherland 1995,1996). However, we point out that not all galaxies
with extended [OIII] emission present extended radio emission (e.g. NGC\,5347,
MCG\,+03-45-003), indicating that
shocks probably are not an extremely important mechanism of ionization in these
galaxies. Finally, the radio jet could just be a passive indicator of the
axis of a shadowing structure determining the [OIII] orientation.

The comparison between the direction of the [OIII] emission to that of the
host galaxy major axis shows a random distribution. This confirms the result
that we obtained comparing the radio and host galaxy major axes (Kinney et al.
2000), and, combined with the good alignment of the radio and [OIII] emission,
indicate that the torus and accretion disk axes are relatively well aligned.
These results imply that the misalignment of the accretion disk relative to
the host galaxy disk cannot be due to the Bardeen-Petterson effect in a
rapidly rotating black hole, but rather that the orientation of the gas in the
torus determines the orientation of the accretion disk.

The conclusions that we can take from our analysis are the following. We find
that our results are broadly consistent with predictions from the Unified
Model. Albeit both Seyfert types have similar NLR size distributions, we take
this result not as evidence that the Unified Model is wrong, but rather that
observed structures can be more complex than expected, and
the models with which we compared our results used simple assumptions.
More advanced models, ones which consider that the galaxies have a range of
luminosities and torus opening angles, are clearly needed. The NLR
size-luminosity relation shows that, to first order, a central source
ionizing the gas is a good representation to the observations, although more
complex models may produce similar results. This agrees with previous results
which show that single zone photoionization models are not appropriate to
study NLR's. The [OIII] and radio emission are well correlated, which indicates
that the torus and accretion disk axes are closely aligned. This implies that
the orientation of the accretion disk is determined by the torus and not by
the spin of the black hole.

\acknowledgements

We would like to thank Jim Ulvestad and Tracy Clarke for discussions.
This work was partially supported by the NASA grants HST-GO-8598.07-A and
AR-8383.01-97A. The National Radio Astronomy Observatory is a facility of the
National Science Foundation, operated under cooperative agreement by Associated
Universities, Inc. This research made use of the NASA/IPAC Extragalactic
Database (NED), which is operated by the Jet Propulsion Laboratory, Caltech,
under contract with NASA. J.L.D. work was supported by the National Science
Foundation, through its Research Experience for Undergraduates program.
J. E. P. is grateful for continued support from the STScI visitor program.

\clearpage
\begin{figure}
\epsscale{0.5}
\plotone{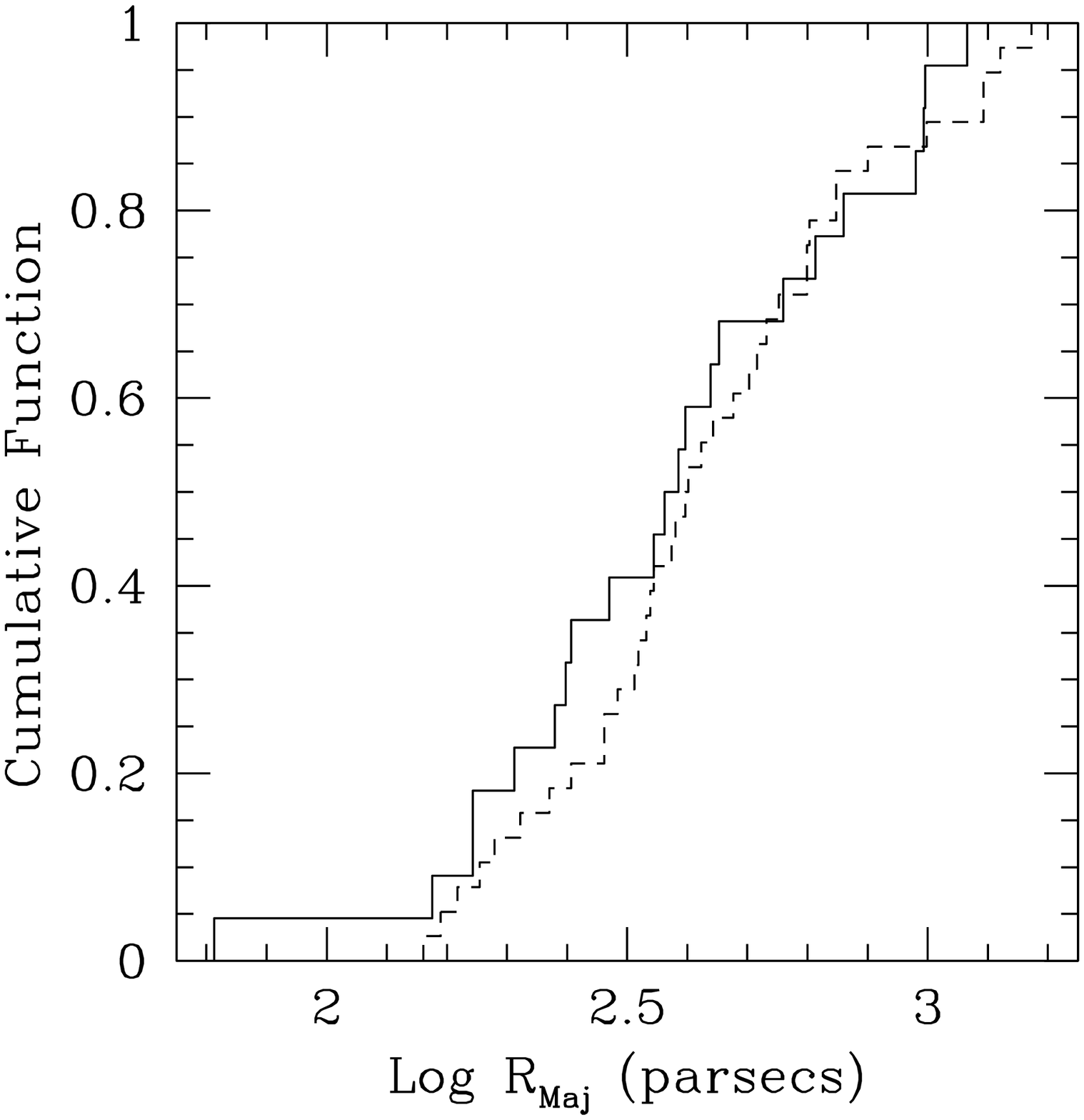}
\caption{Cumulative distribution of the logarithm of the [OIII] emission line
region major axis radius. The solid line represents Seyfert 1's and the dashed
one Seyfert 2's.}
\end{figure}

\begin{figure}
\epsscale{0.5}
\plotone{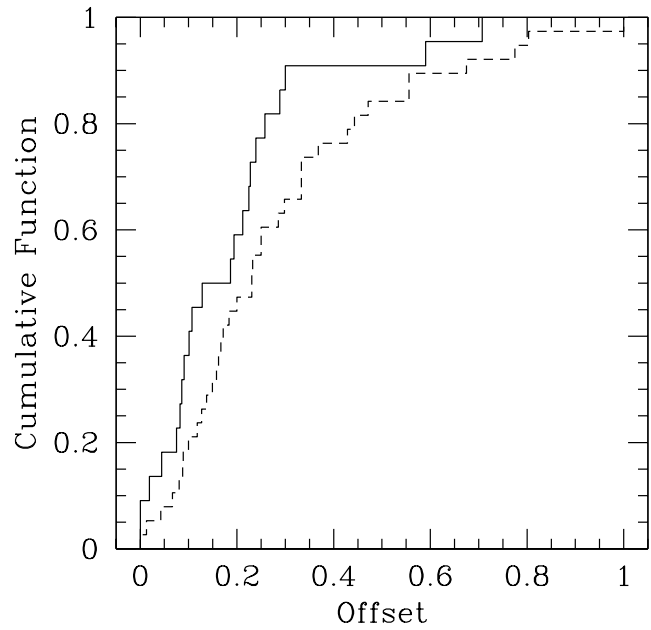}
\caption{Same as Figure 1 for the Offset between the centroid of the
NLR and the peak of continuum.}
\end{figure}

\begin{figure}
\epsscale{0.5}
\plotone{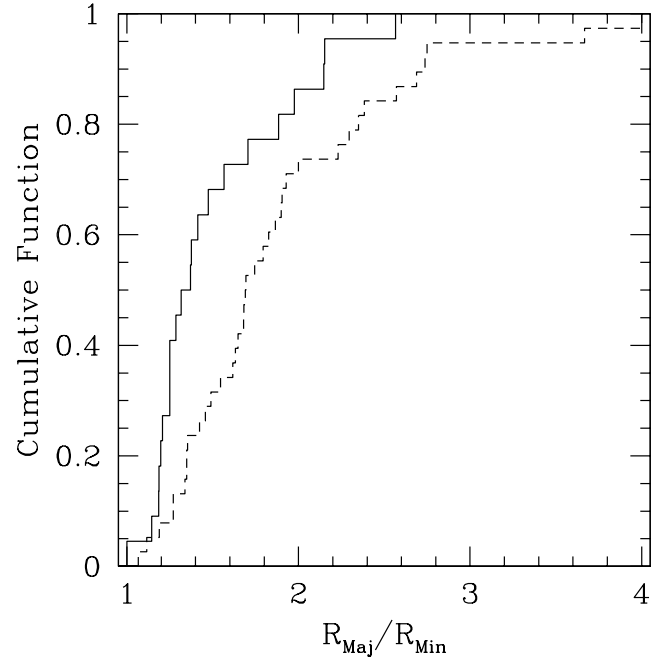}
\caption{Same as Figure 1 for the ratio of the major to the minor axis of
the [OIII] emission line region.}
\end{figure}

\begin{figure}
\epsscale{1}
\plottwo{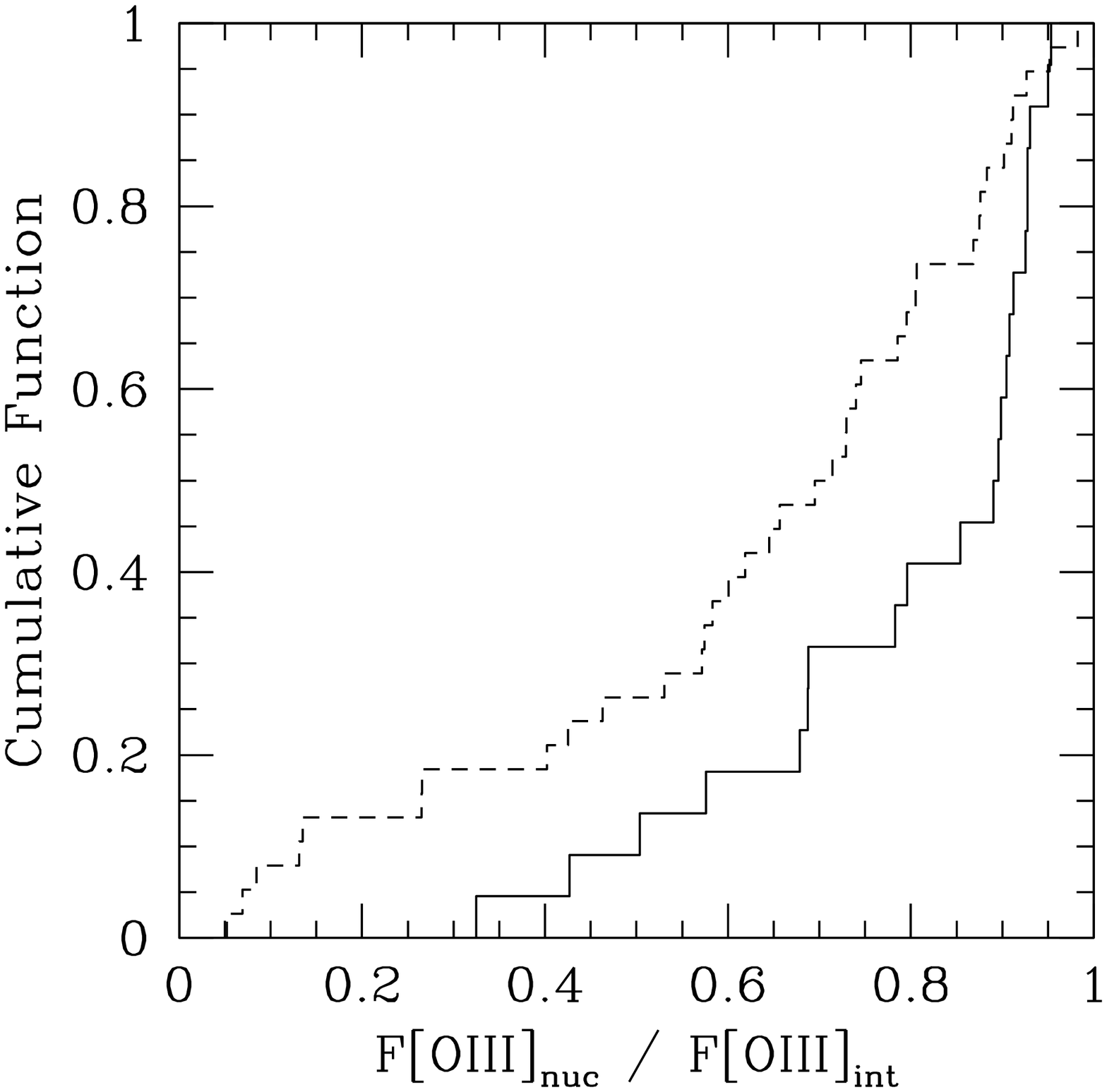}{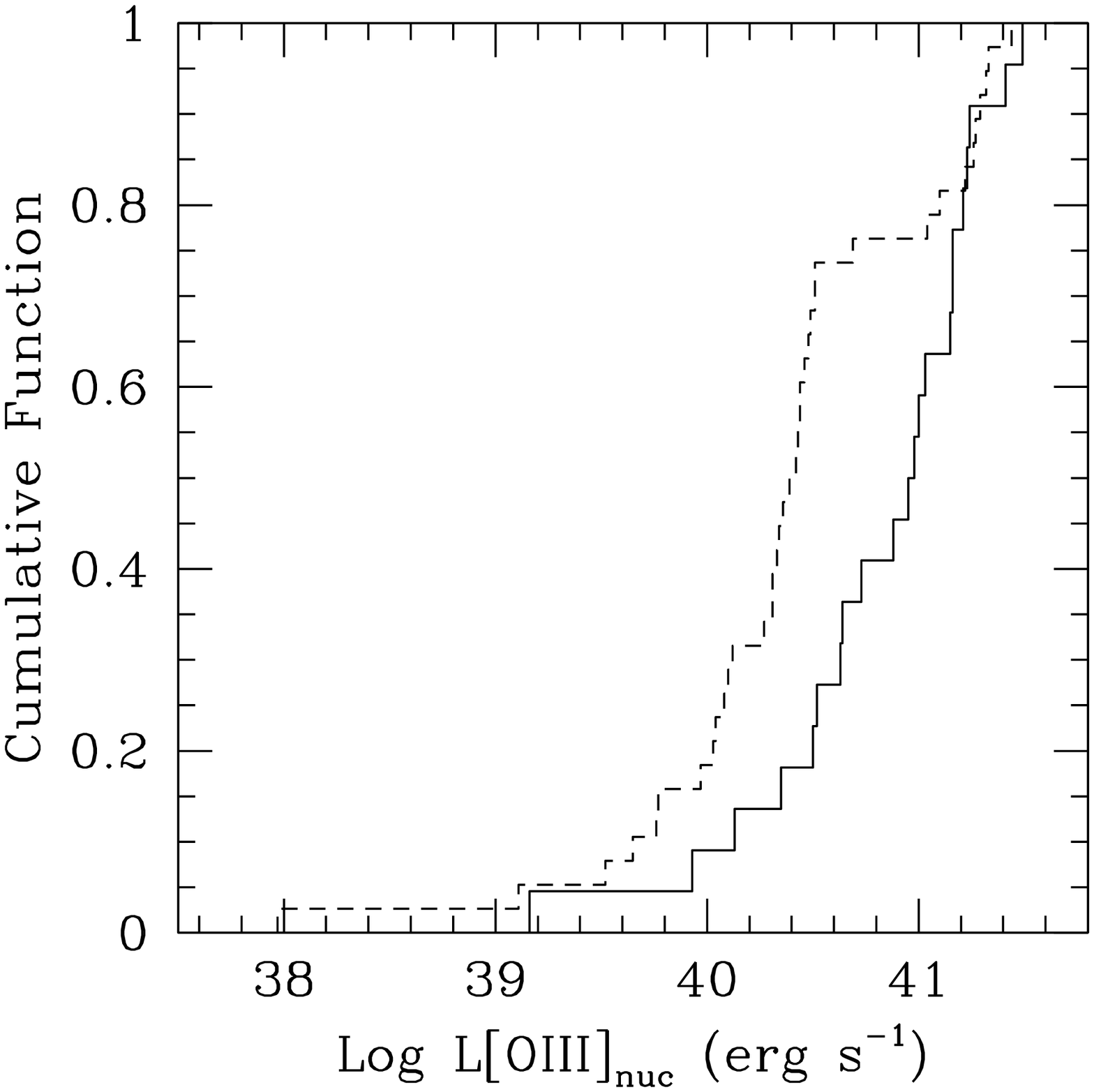}
\caption{Same as Figure 1 for the ratio between the nuclear [OIII] flux,
measured inside a region of 100 parsec radius centered at the nucleus,
to the integrated [OIII] flux (left) and the nuclear [OIII] luminosity
(right).}
\end{figure}

\begin{figure}
\plottwo{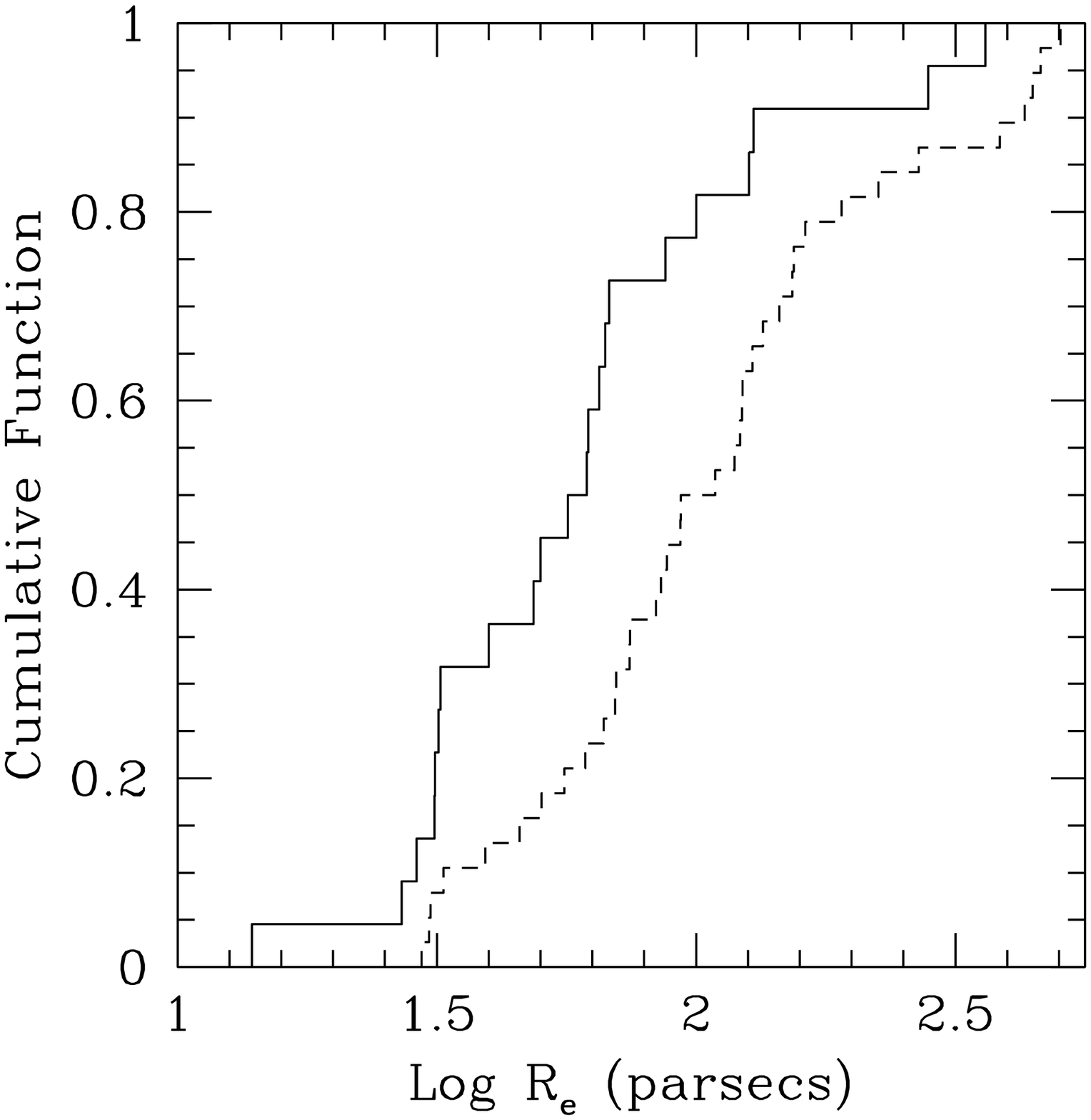}{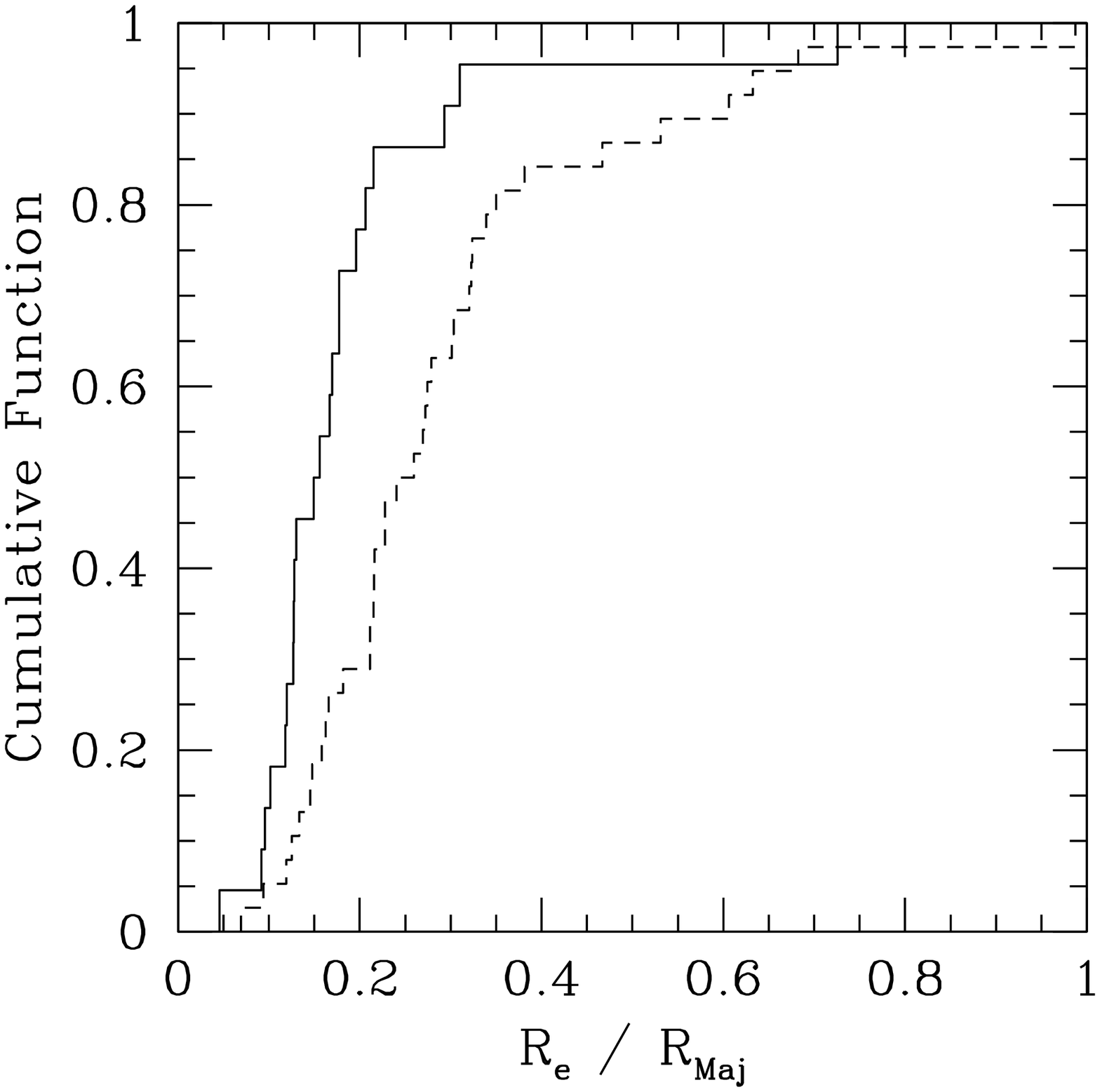}
\caption{Same as Figure 1 for the logarithm of the effective radius (left)
and the ratio between the effective radius and semi major axis radius (right).}
\end{figure}

\begin{figure}
\plotone{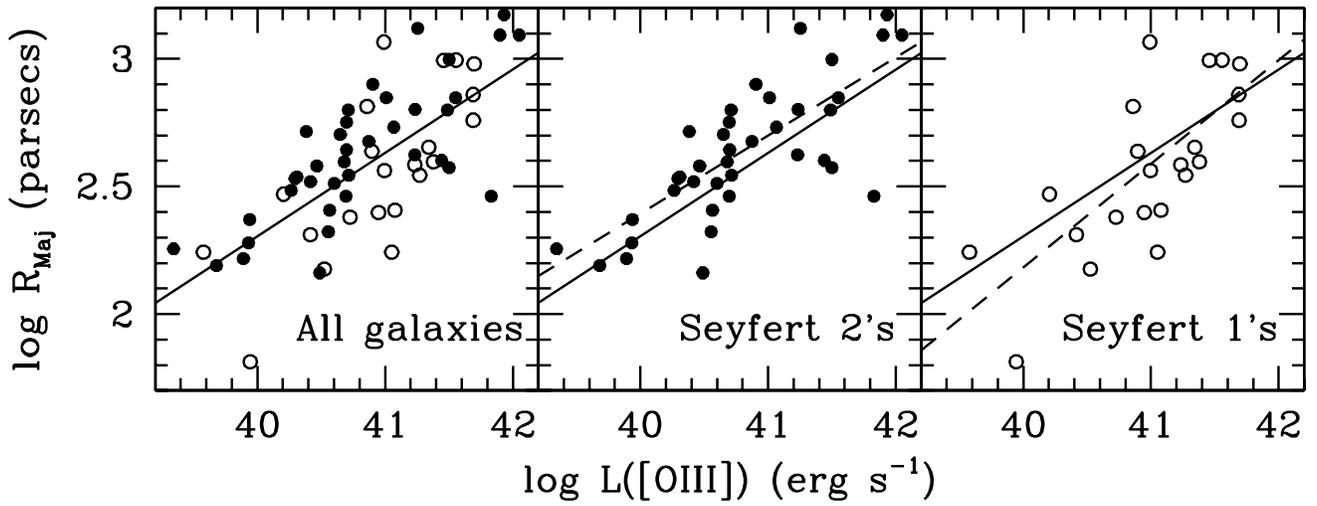}
\caption{log R$_{Maj}$ vs. log L([OIII]). Left: all galaxies in the sample,
Seyfert 1's are the open symbols and Seyfert 2's the filled ones. The line
represents the best fit to the data. The middle and right panels present
only Seyfert 2 and Seyfert 1 galaxies, respectively. The solid line in these
two plots represents the best fit to all the galaxies in the sample (same as
in the left panel), while the dashed line represents the best fit to
the data presented in the panel.}
\end{figure}

\begin{figure}
\epsscale{0.5}
\plotone{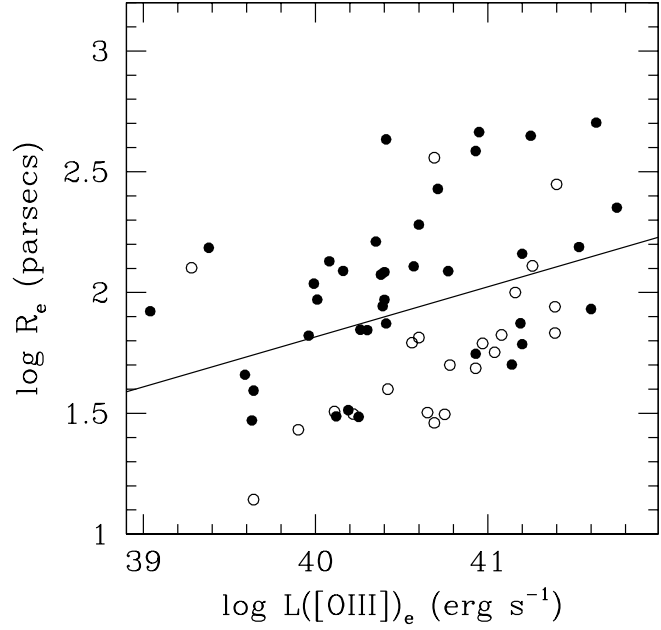}
\caption{log R$_e$ vs. log L([OIII])$_e$ for Seyfert 1's (open circles)
and Seyfert 2's (filled circles). The solid line represents the best fit to the
data.}
\end{figure}

\begin{figure}
\epsscale{0.5}
\plotone{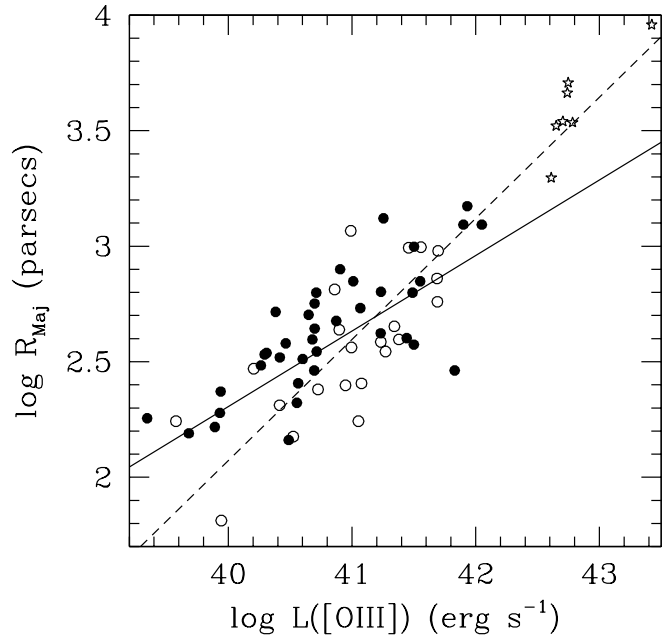}
\caption{log R$_{Maj}$ vs. log L([OIII]) for Seyfert 1's (open circles),
Seyfert 2's (filled circles) and QSOs (stars) obtained from Bennert et al.
(2002). The solid line represent the fit to the Seyfert galaxies, while
the dashed one represents the fit to the Seyferts and QSO's.}
\end{figure}

\begin{figure}
\epsscale{0.5}
\plotone{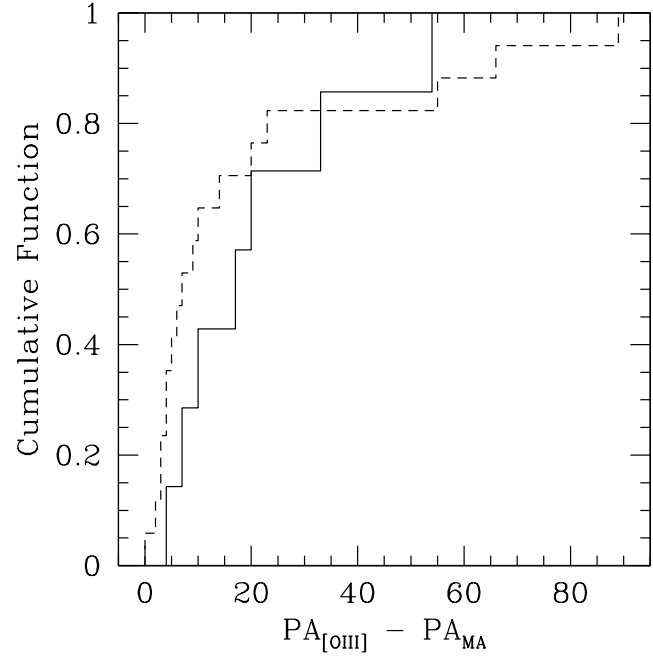}
\caption{Same as Figure 1 for the difference between the P.A. of the major
axis of the [OIII] emission and the P.A. of the radio jet major axis.}
\end{figure}

\begin{figure}
\epsscale{0.5}
\plotone{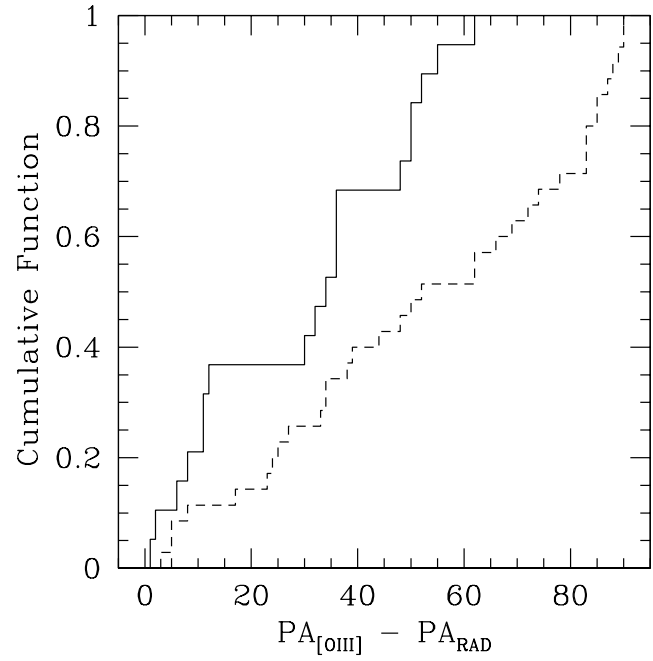}
\caption{Same as Figure 1 for the difference difference between the P.A. of
the major axis of the [OIII] emission and the P.A. of the host galaxy major
axis. In those cases where the NLR has a conical shape, but the major axis has
a different P.A. from that of the cone axis, we use the P.A. of the cone axis.
}
\end{figure}

\clearpage
\begin{deluxetable}{lllrl}
\tabletypesize{\scriptsize}
\tablewidth{0pc}
\tablecaption{Results of Statistical Tests}
\tablehead{\colhead{Property}&
\colhead{Distribution A}&
\colhead{Distribution B}&
\colhead{KS Test}&
\colhead{Figure}\\
\colhead{}&
\colhead{}&
\colhead{}&
\colhead{(\%)}&
\colhead{}\\
\colhead{(1)}&
\colhead{(2)}&
\colhead{(3)}&
\colhead{(4)}&
\colhead{(5)}}
\startdata
R$_{Maj}$&1&2&90&1\\
Offsets&1&2&33&2\\
R$_{Maj}$/R$_{Min}$&1&2&2.3&3\\
R$_{Maj}$/R$_{Min}$&1&2&3.4$^b$&3\\
F([OIII])$_{nuc}$/F([OIII])$_{int}$&1&2&0.47&4, left\\
L([OIII])$_{nuc}$&1&2&0.14&4, right\\
R$_e$&1&2&0.5&5, left\\
R$_e$/R$_{Maj}$&1&2&0.08&5, right\\
P.A.$_{[OIII]}$-P.A.$_{RAD}$&1+2&Uniform&0.01&9\\
P.A.$_{[OIII]}$-P.A.$_{MA}$&1+2&Uniform&56&10\\
P.A.$_{[OIII]}$-P.A.$_{MA}$&1&2&6.2&10\\
\enddata
\tablenotetext{a}{Column 1 gives the Property which is being compared;
Column 2 and 3 give the distributions which are being compared, 1 means
Seyfert 1's, 2 means Seyfert 2's and Uniform means a Uniform distribution;
Column 4 gives the probability that two samples drawn from the same parent
distribution would differ as much as the two samples being compared;
Column 5 gives the Figure where the distribution are presented.}
\tablenotetext{b}{Probability obtained excluding the two Seyfert 2's with the
largest R$_{Maj}$/R$_{Min}$ values.}
\end{deluxetable}

\end{document}